# Blockchain based Decentralized Applications: Technology Review and Development Guidelines

Claudia Pop, Tudor Cioara, Ionut Anghel, Marcel Antal and Ioan Salomie

*Computer Science Department, Technical University of Cluj-Napoca; Memorandumului 28, 400114 Cluj-Napoca, Romania*

claudia.pop@cs.utcluj.ro, tudor.cioara@cs.utcluj.ro, ionut.anghel@cs.utcluj.ro, marcel.antal@cs.utcluj.ro, ioan.salomie@cs.utcluj.ro

**Abstract:** Blockchain or Distributed Ledger Technology is a disruptive technology that provides the infrastructure for developing decentralized applications enabling the implementation of novel business models even in traditionally centralized domains. In the last years it has drawn high interest from the academic community, technology developers and startups thus lots of solutions have been developed to address blockchain technology limitations and the requirements of applications software engineering. In this paper, we provide a comprehensive overview of DLT solutions analyzing the addressed challenges, provided solutions and their usage for developing decentralized applications. Our study reviews over 100 blockchain papers and startup initiatives from which we construct a 3-tier based architecture for decentralized applications and we use it to systematically classify the technology solutions. Protocol and Network Tier solutions address the digital assets registration, transactions, data structure, and privacy and business rules implementation and the creation of peer-to-peer networks, ledger replication, and consensus-based state validation. Scaling Tier solutions address the scalability problems in terms of storage size, transaction throughput, and computational capability. Finally, Federated Tier aggregates integrative solutions across multiple blockchain applications deployments. The paper closes with a discussion on challenges and opportunities for developing decentralized applications by providing a multi-step guideline for decentralizing the design of traditional systems and implementing decentralized applications.

## 1. Introduction

Distributed Ledger Technology (DLT) is a disruptive technology that provides the infrastructure for developing decentralized, secure, and reliable applications. By joining several computer sciences disciplines such as distributed systems, cryptography, data structures or consensus algorithms, it offers highly desirable features (decentralization, openness, immutability, transparency, traceability, security, availability, etc.). Gartner has included DLT in the technology in the hype cycle for the first time in 2016 at the phase of an innovation trigger [119]. During the next two years, a lot of research and development work has been committed to develop the mechanisms to accommodate the technology to the various requirements essential to decentralized applications implementation perspective and emerging business models. By 2018 the DLT has passed the peak of inflated expectations, assuming to reach a plateau of productivity in the next 5 to 10 years [120], while in 2019 the innovation potential of the DLT is considered by Gartner to be driven non only by the technology expectations but also by the social ones [121].

In this sense, the development of Decentralized Autonomous Organizations and implementation of Decentralized Applications and the Decentralized Web are of high interest for the next 10 years. Anyway, building such decentralized applications with DLT is not a straightforward process since lots of technological solutions have emerged or have been significantly improved, offering new opportunities and challenges for the software development industry [1].

Thus, in this paper we aim to provide a comprehensive overview of nowadays blockchain technology solutions and the specific challenges they aim to address and provide a guideline for applying the blockchain technology for decentralizing traditional systems and implementing new decentralized applications.

To streamline and organize the DLT solutions review we have defined a 3-tier conceptual architecture (see Figure 1) that can be the used for implementing blockchain decentralized applications:

- The Protocol and Network Tier (PN-Tier) aggregates the core DLT elements and organizes them in two layers. The Protocol Layer contains technology solutions for digital assets registration, transactions, data structure and privacy and business rules implementation. The Network Layer contains technology solutions for creating a peer-to-peer network, ledger replication and consensus-based state validation.
- The Scaling Tier (S-Tier) runs most of the time a parallel blockchain network and aggregates technological solutions for addressing the scalability issues raised by the PN-Tier (also known as off-chain components). We have further organized the state-of-the-art solutions according to the scalability problems that they address in storage scalability, transactions throughput and computational scalability.
- The Federated Tier built on top of the previous two tiers addresses various integration solutions across multiple DLT applications and systems deployments.
- For each architectural tier we identified the main technological challenges, existing solutions, and technologies to create a consistent overview over the current technological state of the art.

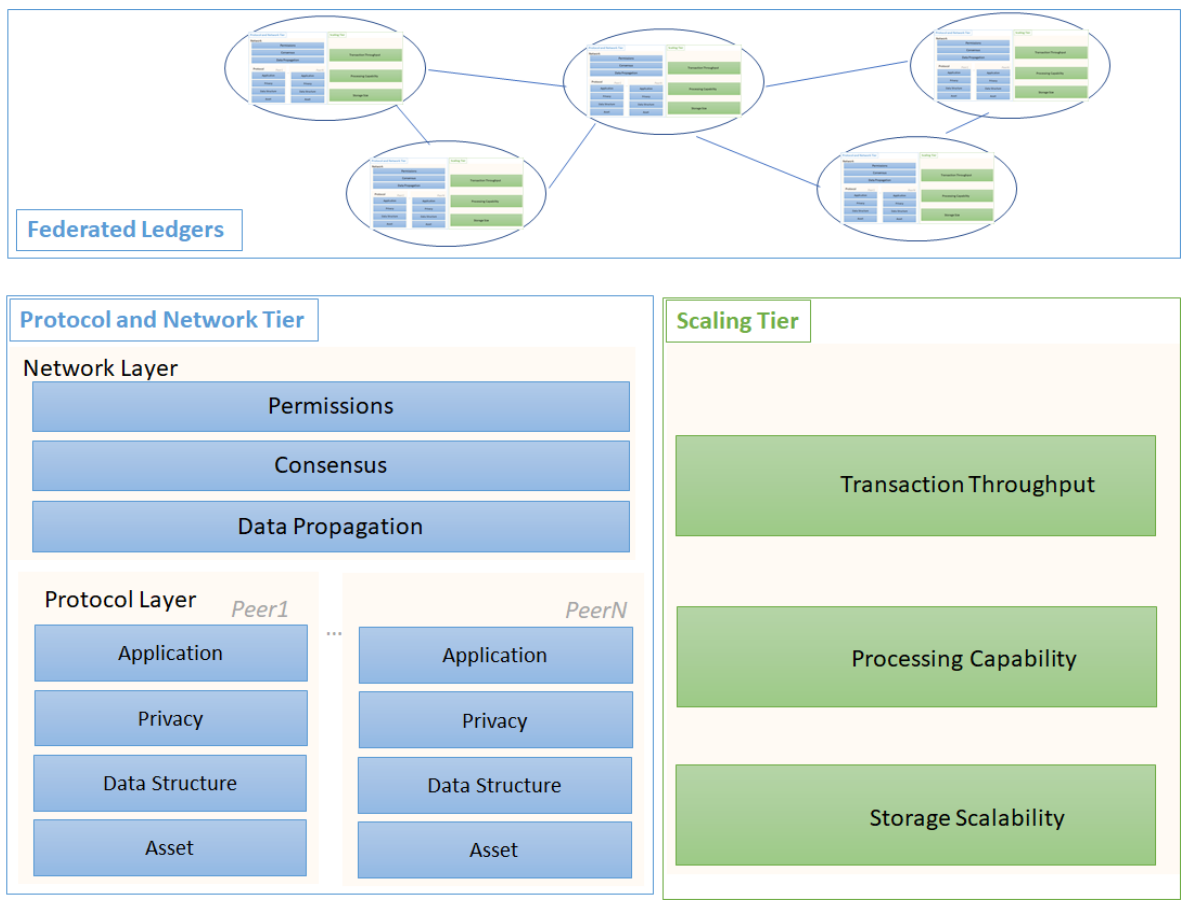

**Figure 1.** 3-tier architecture for decentralized applications development

The rest of the paper is structured as follows: Section 2 presents the main technological solutions for the PN-Tier, Section 3 reviews the mechanisms of S-Tier for improving the PN-Tier's scalability in three directions: storage, transaction throughput, and computational, Section 4 describes the existing federated ledgers interoperation solutions, Section 5 discusses guidelines for developing decentralized applications, while Section 6 concludes the paper.

## 2. Protocol and Network Tier

Different solutions for PN-Tier have been proposed addressing specific challenges [1-6]. For this tier we have identified the technological components that are grouped at the protocol and at the network layers, while the security of the entire tier is ensured as a result of integrating the public-private key cryptography for locking and unlocking transactions, with the tamper-proof data structures that are enforced by strong consensus algorithms.

## 2.1. Protocol Layer

The protocol layer represents the core of the technology runs on each full node in the peer-to-peer network. It is governed by rules that specifies what, when, how and by whom are the assets operated on the chain and features four type of technological components: asset representation, data structure, privacy and business rules enforcement.

*2.1.1. Type of asset and data structures*

The tokenization process refers to the possibility of modeling different goods in a DLT system as digital assets that can be issued and transferred according to a predefined set of rules. The common terminology for an asset representation in DLT systems is Token or Coin. We have identified two types of tokens that can be represented in the system (see Table 1): native tokens and tokens based on real assets (asset-based tokens). Native tokens are tokens defined in the blockchain system, completely independent of the real world, thus the rules governing the issuance and the transfer are completely defined in the blockchain system (through Initial Coin Offerings or mining reward schemes), and do not rely on any third trusted party. Bitcoin [1], Ether [2] or CryptoKitties [7] are examples of such tokens that are completely virtual and have economic value based on the supply and demand.

**Table 1.** Type of digital assets modeled using blockchain

| Type | Token Example | Fungibility | Issuers |
|---|---|---|---|
| Native tokens | Bitcoin, Ether, CryptoKitties [1,2,7] | Yes | Mining Reward Schemes |
| | ERC20, ERC223, ERC-621 [8] | Yes | Initial Coin Offerings |
| | ERC721 [8] | No | Initial Coin Offerings |
| Asset-Based Tokens | Real Estate [9] | No | Government Land Registries [10] |
| | Patents [11] | No | U.S. Patent & Trademark Office [12] |
| | Academic Records [13] | No | The Registrar's Office [14] |
| | Gold | Yes | Royal Mint Gold [15] |

Real-life assets can also be represented through tokens in blockchain systems, offering the opportunity to represent, transfer and track them. By using asset-based tokens, the chain can keep track of different kinds of assets, both tangible (real estate, cars, money, art, etc.) and intangible (patents, trademarks, copyrights, etc.). The DLT systems representing real-life assets must rely on a trusted third party in order to issue each token with respect to the real object. Similarly, a transfer on-chain must be done under the governance of such a trusted party, since any issue regarding a wrongful transfer on the chain can be verified only by communicating with the external systems.

In DLT systems the real-life flow of assets is represented in the network as transactions between peers. The DLT requires specialized data structures at its core that can ensure three properties over the stored transactions: provenance, asset ownership validation and immutability. Hash pointers have been frequently integrated with different data structures intended for DLT usages. Due to the hash functions' collision-free, data concealing and data binding properties, the hash pointers are the best choice for adapting common data structures to the DLT requirements thus obtaining: linked list with hash pointers (e.g. blockchain), binary trees with hash pointers (e.g. Merkle Trees), graphs with hash pointers (e.g. Hash Graphs), etc.

Two main state the art directions can be identified concerning distributed ledger data structures for representing peers' transactions, namely block chain and direct acyclic graph. The block chain structure, as its name suggests, is a chain formed by linked back blocks, also known as Linked List using hash pointers (see Figure 2.a). Each block contains all the transactions that occurred in the system in a short period of time (e.g. ~10 minutes for Bitcoin, or ~12 seconds for Ethereum). All the transactions contained in the block are hashed together in a Merkle Tree data structure, where the root of the tree is referenced in the block header and acts as a digital fingerprint of the entire collection.

Thus, blockchain becomes an append-only data structure that gathers all the benefits of the hashing and cryptographic functions and, with the integration of a consensus algorithm, ensures an immutable history log of the entire activity of the network. The blockchain structure is implemented in well-known solutions such as Bitcoin [1], Ethereum [2], Litecoin [4], Hyperledger [16], CryptoNode [17], Ripple [18] and Zerocash [6].

A less popular data structure is the directed acyclic graph (DAG) firstly mentioned in [19]. Since its proposal several platforms have been developing solutions based on DAG variations: Dagcoin [20], IOTA [3], HashGraph [21], Lydian [22], or hybrid systems like Holochain [23] and Flowchain [24]. Among the DAG-based systems, IOTA is the most used solution. It uses a DAG, called tangle, as a ledger for storing the transactions as depicted in Figure 2.b. The entire graph starts with a genesis transaction that is approved directly or indirectly by all the transactions in the graph. Whenever a new transaction is submitted, it must validate and confirm two previous transactions from the graph that were not yet approved (i.e. tips). The tips selection algorithm is based on the family of Markov Chain Monte Carlo algorithms and it considers the cumulative weights of sub-tangles. Whenever a situation of conflicting transactions appears, the higher the cumulative weight of the transaction is, the more secure it is.

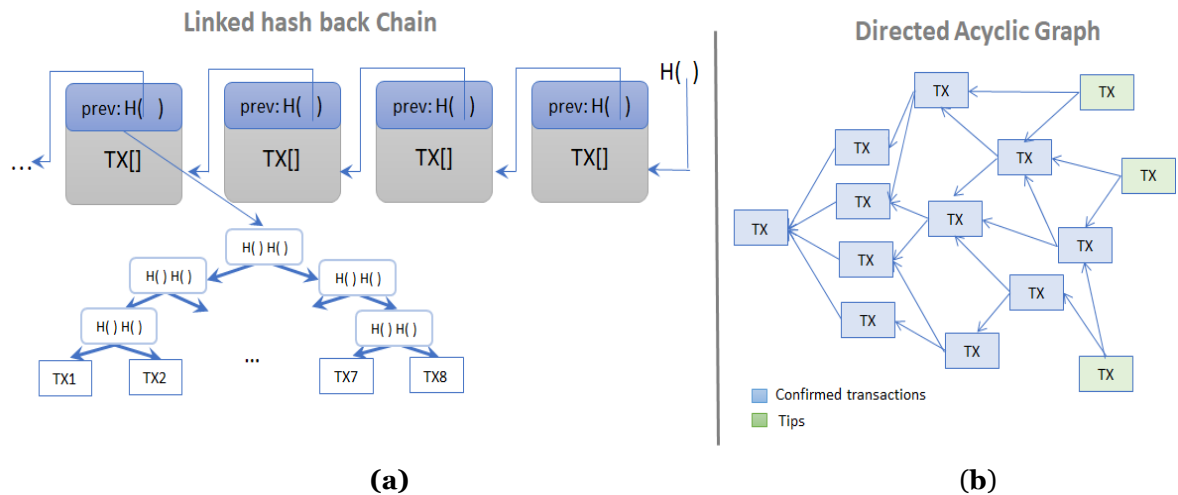

**(a)** **(b)**

**Figure 2.** Data for representing peers' transactions structures (a) block chain using linked lists and (b) Directed Acyclic Graphs

While the most successful solutions and research directions are focused on blockchain-based DLT, the DAG solutions have yet to overcome considerable shortcomings in term of centralization and scaling in order to be considered suitable alternatives to blockchain-based DLTs [25].

*2.1.2. Privacy of transacting parties and data*

Most of the popular DLT solutions preach the properties of transparency and openness to be major benefits brought by the technology. By providing open information about the transacting parties (pseudo-anonymous most of the times) and transacted assets, the systems offer clear history and audit advantages. However, in many use cases where the privacy of data is highly required [27] (e.g. medical use cases [27] [29]), many of the DLT solutions may prove to be unsuitable due to their transparency and openness. With the emerging General Data Protection Regulation (GDPR) restrictions, the privacy of data is one of the most controversial subjects in the DLT systems. In this direction, solutions have emerged that aim to hide the details regarding the transaction information, while at the same time keeping the reliability of the system by allowing consistency checks, validation, and audit with respect to the previous actions and history.

In Bitcoin versions of the DLT, the value of the transacted asset is clearly stated, and the nodes can easily check the ownership over the asset by checking the history and identifying the unspent transaction outputs (UTXO). The biggest challenge is that by hiding the transaction data, the nodes in the system should still be able to validate the availability of the funds and the ownership of the sender over that asset. Coin Mixers have been developed over Bitcoin in order to hide information and prevent tracking and tracing of transacted assets. One of these approaches is CoinJoin [30] which aims to aggregate multiple users to agree upon several inputs and outputs and then sign the transactions. This makes it harder for the coins to be traced across the transactions since the group of inputs provided to a transaction will not belong to the actual sender. Similarly, Dash [31] aims to provide mixing services but instead of using a central point that is responsible for mixing inputs and outputs like CoinJoin, it provides a second layer of Master Nodes over Bitcoin.

A private-public key cryptographic mechanism is proposed by Quorum [29] in order to hide the transacted value. The encrypted transactions are shared point-to-point only to the involved parties, and a Private State is defined by the Quorum node, where these encrypted transactions are stored. The encryption keys are shared between the involved parties, to validate the transaction. However, the shared blockchain (public state) only contains the hash of the encrypted transaction. This leads to a shortcoming of the system since the network cannot verify the validity of the private transactions, this being done only by the involved parties. The shortcoming of Quorum is overcome by integrating Zero-Knowledge proof mechanisms [32] with DLT systems. Zero-knowledge proofs have been added to digital currencies, such as Zerocoin [33, 6] to create anonymous Bitcoin transactions without the need of third parties such as CoinJoin and Dash. The Zero-Knowledge proof mechanisms, mainly succinct Non-Interactive Zero-Knowledge proofs (ZkSnarks) [34] are used to ensure the transactions' validation without revealing actual information about the parties involved. The Zero-Knowledge proofs require that a Verifier could easily check that the Prover is in possession of a secret, without revealing the actual secret to the Verifier. Consider a simple scenario, where the network is in possession of a hash value H. An actor wants to prove to the network that he holds the secret s, that hashed offers the value of H. In the traditional system, the actor would need to reveal the secret to the network. ZkSnarks aim to enhance this model, by allowing the actor to prove ownership over the data without disclosing any information regarding the secret. In this sense, ZkSnarks introduces a proving function, that can be used by the actor to issue a proof showing that the private secret is indeed corresponding to the public information H, and a verification function that can be executed by any participant in the network in order to validate whether the proof corresponds to the public information H. ZCash [6] implements a ZkSnarks mechanism as an improvement of the Bitcoin system that ensures the privacy of the transactions. It requires any transaction to be locked by a secret, called a commitment and unlocked by a participant that holds the secret and who can generate the relevant proof, called the nullifier. The commitment is issued off-chain by the sender of the transaction, having as secret information the value and the receiver's public key. Similarly, the nullifier is computed off-chain by the receiver, proving that he owns the necessary information to spend the locked value. The commitment and the nullifier are registered on-chain. The commitments are registered in the commitment tree, and each time a transfer is required, a nullifier for one of these commitments needs to be issued and then registered in the nullifier set as future proof for avoiding double-spending attacks. The verifiers of the nullifier are all mining nodes that need to validate the integrity of the transactions.

Homomorphic encryption is another approach that aims to improve the privacy of blockchain solution in MimbleWimble [35], a system designed to provide an untraceable version of Bitcoin. It is a side chain that takes advantage of cryptographic properties to hide transacted values. In Bitcoin, each transaction is represented by an input amount that has an associated past UTXO that it unlocks, and an output amount locked by the receiver's key. The restriction imposed by the Bitcoin system is that the output amount of a transaction should never exceed the input amount of it. MimbleWimble uses Elliptic Curve Cryptography and Homomorphic encryption, leveraging on the fact the computations applied on the cyphertexts offer the same results as if applied on the plaintext. Therefore, in MimbleWimble the transacted amounts are blinded and by applying computations on the obtained cyphertexts they are further involved in mathematical operations proving that the input values of the transaction equal with the output values, obtaining the same results as they would have been performed directly in plaintext.

Another commonly used strategy, that hides the transferred asset amount of a transaction and the parties involved, is the ring signature. CryptoNote [36] is one of the first protocols that proposes the use of ring signatures for issuing transfers by specifying a group of possible signers in order to avoid the possibility of discovering the exact sender of the money. Considering a group of N parties (sub-group of network participants) involved in the ring, one of the parties can sign the message, resulting in a signature that can be verified by anyone in the network, but without the possibility of detecting the exact signing party. Furthermore, in Monero [5], the authors use Stealth addresses for the receiving parties. Each Monero account is composed of two private keys (view and spend key) and the public address. As their name suggests, the view key is used to track all the transactions that were published and are destined for that account. The spend key is used to send transactions and the public address is the one used by a sender to compute the one-time public key (stealth address), unique for each transaction. The stealth addresses offer non-linkable transactions, which means that the outputs are not associated with the addresses of the wallets. However, this solution does not provide complete privacy, since the sender can trace when the money is spent by the receiver. A completely private system would need to offer both non-linkable and non-traceable transactions.

Table 2 presents comparatively the main privacy preserving techniques identified for DLT and the privacy features they are offering. The coin mixers do not offer complete privacy, although they offer non-traceability mechanisms, the actual values transmitted are still visible. In terms of non-traceability however, the Zero-Knowledge Proofs were not yet validated as completely untraceable, due to the complexity of the algorithms involved. ZCash aims to provide traceable capabilities for the system, in order to be able to detect malicious users, which leads to the conclusion the transactions may not be completely untraceable [37]. Private-Public Key cryptography and Coin mixers also have one main disadvantage, because they rely on central authorities in order to provide Key sharing services and mixing services respectively. Another important property is the advanced scripting capability, where Coin mixers and Homomorphic Encryption mechanisms prove not to be a good solution, while the Zero-Knowledge Proof solution although feasible, has the drawback of using high computational resources for applying the cryptographic algorithms.

**Table 2.** DLT main privacy presenting techniques

| Features | Private-Public Key Encryption | Zero-Knowledge Proofs | Ring signatures | Homomorphic Encryption | Coin mixers |
|---|---|---|---|---|---|
| Hidden Data | yes | yes | yes | yes | no |
| Non-traceable | yes | n/a | yes | yes | yes |
| Non-linkable | no | yes | yes | no | yes |
| Decentralized | no | yes | yes | yes | no |
| Private Business Enforcement | no | yes | no | no | no |
| Transaction validation by network | no | yes | yes | yes | yes |

*2.1.3. Business rules enforcement over transactions*

The capability of a blockchain system to support business implementation that can be run in a decentralized way, and then be verified and audited by all the nodes in the system, is usually provided through smart contracts. Opposed to the concept suggested by their names, the smart contracts are not very smart and may not provide a contract in the legal sense, but rather they are pieces of code similar with the stored procedures that are executed and validated by the nodes in the system, whenever they are triggered. However, there are DLT systems such Bitcoin that are offering few possibilities for scripts to be implemented. Not being a Turing Complete language, the Bitcoin script language does not permit the implementation of complex business logic required for more advanced use cases. As a result, new DLT systems have emerged that allow customizing decentralized applications and enforcement of business logic in a decentralized way regarding when, how and by whom may an asset transfer be executed.

In Table 3 a comparison of the main state of the art approaches for enforcing the business rules on DLT are presented. There are two types of approaches that allow smart functionality to be implemented and run across the nodes of the network: stateless and stateful [38]. The Stateless implementation or Transaction based functionalities are offering the possibility to implement custom logic at the level of transactions. This means that whenever a transaction is issued there is a set of rules that can be verified before rendering the transaction valid. The Stateful systems, on the other hand, offer Business-oriented functionalities, by focusing on the rules that govern the use case, and keeping the state of the business in tamper-proof structures (Patricia Merkle Trees, adapted from [39]) that are easily verifiable and audited by the entire distributed system. In the Stateful Systems, the transaction has the role of triggering changes and applying updates on the stored state.

**Table 3.** Business rules enforcement on DLT

| Type | Implementation | Platform | Enforcement Flexibility | Costs | Exploitation Risks |
|---|---|---|---|---|---|
| | | **Transactional Rules** | | | |
| Stateless Transaction Oriented | Built-In Enforcement | Bitcoin [1], Litecoin [4] | Limited | None | Low |
| | | Nxt [40] | Templates | None | Low |
| | Piggy Backed Enforcement | Counterparty [41] | Turing Complete | Fee per instruction | High |
| Stateful | | **State Storage** | | | |
| Business Oriented | Smart Contracts & Merkle Patricia Tree | Ethereum [2] | Turing Complete | Fee per instruction | High |
| | Smart Contracts & NoSQL DB | HyperLedger [16] | Turing Complete | None | High |

In terms of functional complexity, there are systems that allow full computation capabilities by supporting Turing Complete languages for smart contract implementations or partial capabilities by offering a limited range of operations or fixed predefined templates. Both approaches have their advantages. On one hand, full capabilities are desired in order to be able to model and enforce any complex business system. Turing Completeness allows this, but it requires higher transactional costs, since on-chain computation demands for each node to execute possibly complex scripts, thus the costs are proportional with the number of instructions. Furthermore, a complex and flexible language makes the system susceptible to different kinds of attacks that can be caused by exploiting different language shortcomings or human errors during the business logic implementations [42] like call depth attack, race conditions, timestamp dependency, transaction ordering dependency, etc. On the other hand, by limiting the operations allowed or by providing predefined templates to be used, the risk of exploits is highly reduced.

**2.2. Network Layer**

The network layer technologies are related to the peer-to-peer network formed by the nodes that hold copies of the ledger (full nodes or light nodes) and participate as active players in the network. The main technological components identified at this layer are targeting the data propagation among the nodes, the peer's registration and network permission, and the consensus among peers.

*2.2.1. Data propagation and replication*

In terms of transaction data propagation, the first generation of blockchain systems (Bitcoin [1], Litecoin [4], Ethereum [2], etc.) relied on full-discovery or global disclosure. This is one of the strongest features of blockchain systems since a complete replication of the data offers high availability and reliability in the system. However, there are specific use cases (e.g. banking, enterprise data) that impose restrictions regarding the group of actors that can have access to transaction information. In this sense, two categories of systems have been identified based on how the transactions are propagated in the system. Firstly, the global disclosure mechanism, implemented by the systems where all the full nodes have access to all the transactions published in the system, and secondly, the selective disclosure mechanism where nodes have access only to exclusive transactions that are targeting either specific businesses or only the involved parties.

Most of the blockchain-based system adopts a global disclosure approach in order to offer high reliability in an open system where any node can join. The entire system is a peer-to-peer network, where all the nodes are equal. Whenever a new event is issued (a new transaction, a new block) the data is propagated through the entire network, and each node can verify and validate the integrity of the data. The redundancy in storage and computation makes it very difficult for a malicious node to influence the system to its advantage. In order to attempt an attack (e.g. double-spending attack) on a globally disclosed DLT system, an elaborate plan of attack must be conducted by the malicious node by analyzing the network topology (network segmentation) and issuing contradictory actions for each half of the network, with the purpose of convincing half of the network to agree with the malicious action taken by it.

Having a global disclosure between all the peers in the network has obvious advantages since such a system benefits from the high replication and availability brought by the large number of nodes, as well as Byzantine Fault Tolerant consensus between these nodes regarding the data. However, there are clients/businesses that prefer having more privacy and control over their data. This property is especially desired in private and consortium chains (e.g. banking systems), where the transactions are required to be shared only between the transacting parties. Although such a paradigm shift may lead to lower reliability in the system, the risks are highly attenuated if these requirements are implemented in permissioned systems where each stakeholder has its identity known and can be held accountable for his actions.

One of the selective disclosure approaches is presented in the Hyperledger Multichannel Architecture [43]. The system relies on third-party entities, called Orderers, which are required to order the transactions and publish them according to the category (business specific) in a corresponding channel. A Byzantine fault-tolerant consensus protocol is implemented between the Orderers, to ensure consistency between the decisions. A channel is a business-specific queue that broadcasts all the transactions to the subscribed parties. All the subscribers (peers) will receive the transactions in the same order in cryptographically linked blocks. A peer can be subscribed to more than one chain, but the chains do not interact with each other and each block received will contain only transactions corresponding to the corresponding business.

Quorum [29] is another approach that aims to improve security by keeping the exclusive transactions shared only between the involved parties. The system is a hybrid between the global and selective disclosure paradigms, by allowing public transactions to be fully replicated and exclusive transactions to be shared only across the parties. The Quorum's privacy engine defines a private state tree which is updated with contracts and transactions that are sent point-to-point only to the interested parties. The private transaction contents are encrypted using Public Key cryptography, and only the users holding the private keys have access and can decrypt the actual content of the transaction. Proof of these events is also registered in the public chain, by hashing the encrypted private transaction.

A similar permissioned implementation is also designed in Corda [44] where the network is formed of permission services, notary services and peers. The aim of the system is to provide redundancy while also keeping the transactions only known to the involving parts. Any transaction that occurs in the system must be signed and approved by both participants, and by the notary service responsible to validate transactions and prevent double-spending events. The notary service can be one entity or multiple entities that are coordinated by a consensus algorithm.

In Table 4, the comparison between the Data propagation patterns found in the literature is presented. One of the biggest disadvantages of the current selective disclosure systems is their trust in different central authorities. The Quorum system requires some level of trust between the private parties, and the other systems rely on central authorities that are responsible either for forwarding the messages like in the case of Hyperledger MultiChannel system or on authorities responsible to validate the integrity of transactions like in the case of Corda or Plasma [45]. Consequently, the selective disclosure should be considered only in trusted environments, where the central authorities can be considered a source of truth, while for public environments, global disclosure should be considered such that any party involved in the network can validate the integrity of the transactions.

Table 4. DLT data propagation patterns

| Type | Platform | Trusted Parties | Global Disclosed | | Selective Disclosed | |
|---|---|---|---|---|---|---|
| | | | Data | Structure | Data | Structure |
| Public DLTs | Ethereum [2], Bitcoin [1], etc. | - | All transactions | Blockchain | - | - |
| Business Specific Chains | HyperLedger Multi-channel [43] | Orderer | - | - | Exclusive transactions | Queues, Blockchain |
| | Plasma [45] | Central Authority, N delegates | Public Transactions + settlements | Blockchain | Exclusive transactions | Blockchain |
| | Corda [44] | Notary Service | - | - | Exclusive transactions | Local database |
| Point-to-Point transactions | Quorum [29] | Private parties | Public transactions, Hashes of Exclusive Transactions | Blockchain | Exclusive transactions | Merkle Patricia Tree |

*2.2.2. Permission mechanisms*

Over the years the private institutions that realized the potential of the systems behind DLT, started to evaluate the integration of such systems with their businesses. However, some key components rendered the public chain unsuitable for many institutional and enterprise solution requirements, so they started to investigate new systems that address the issues regarding the governance and the permissions of the system. Firstly, in an enterprise solution, it is highly important for the participants to be known and vetted before given access. Such a decision has a great impact on the system, even in terms of security and consensus. Since the participants are known, thus can be held accountable for their actions, the need for a high energy-consuming algorithm like Proof-of-Work is no longer justified. Therefore, there is a strong relationship between the requirements regarding the access rights and the consensus algorithms suitable for a specific business.

The difference between public, private, permission-less and permissioned blockchain is given mainly by the rights of the users in the system. Based on the classification presented in Table 5, the difference between private and public chains is established according to the target audience that has access (reading rights) on the chain. Restricting the access of a group to the chain renders the chain private. According to the group of people accessing the chain, it can be a consortium or an enterprise solution, where the consortium solution operates under the leadership of a group of companies, and the enterprise solution is under the operation of a single entity.

**Table 5.** Public vs Private Blockchain permissions.

| Action | Public Chain | | Private Chain | |
|---|---|---|---|---|
| | Permission-less | Permissioned | Consortium | Enterprise |
| Chain Access | Everyone | Everyone | Group Owner | Group Owner |
| Transactions | Everyone | Owners & Validated Users | Owners & Validated Users | Administrator |
| Commit to chain | Everyone | Owners & subset of Validated Users | Owners & subset of Validated Users | Administrator |

In public DLTs some restrictions can also be imposed regarding the users' access and permissions. In a permissioned ecosystem, the validators are known and accountable for their actions, thus a certain level of trust between the nodes can be considered. In a permission-less system, on the other hand, any user can perform any type of action (transactions of an asset, as well as commits of new blocks to the chain). Consequently, permission-less DLTs require Byzantine Fault Tolerant consensus algorithms, since the openness of the system allows even malicious nodes to join, making the network susceptible to a larger range of attacks.

*2.2.3. Consensus protocols*

Figure 3 presents a taxonomy of the consensus algorithms which are classified in Non-Byzantine fault tolerant algorithms and Byzantine fault tolerant algorithms. The difference is given by the ability of algorithms to reach agreement, integrity and termination in case of existing faulty or attacker nodes in the distributed system, thus Non-Byzantine fault tolerant ones rely on the assumption that all the nodes are fair, while the Byzantine fault tolerant algorithms can handle situations when the number of malicious nodes is as high as half of the total number of nodes.

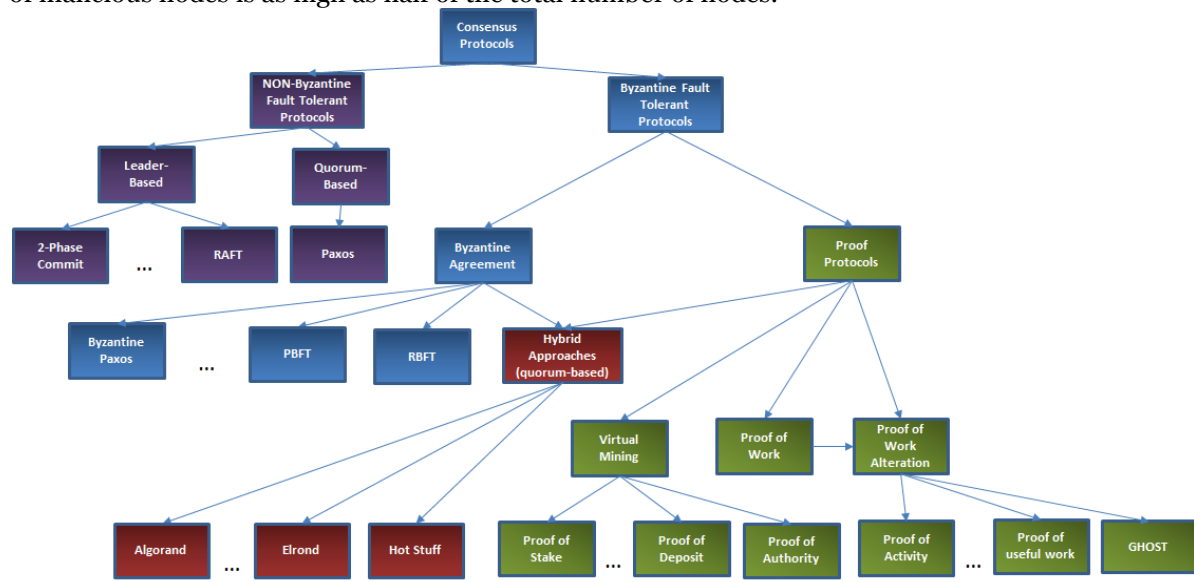

**Figure 3.** Consensus algorithms taxonomy

The Non-Byzantine fault tolerant protocols are leader-based, such as 2-Phase Commit [101] and RAFT [102], where a leader election algorithm is used to select a leader that will centralize the votes and commit the transaction. Furthermore, they are quorum based, where a subset of the processes is selected to validate the transaction using a voting scheme. A well-known algorithm of this class is the Paxos algorithm [103] that solves consensus in a network of processes that may fail but are correct (there exist no faulty processes that may lie).

The Byzantine fault tolerant protocols aim to assure that the peers are be able to agree on a system valid state even in case some of them feature faulty or malicious behaviors. The idea is to find a model and protocol for a network of message-passing processes, some of them being faulty, such that a general agreed state can be extracted from the distributed system. The Byzantine fault tolerant protocols can be classified as traditional Byzantine Agreement protocols and Proof Protocols [104]. The Byzantine Agreement protocols use a quorum-based mechanism where a subset of the nodes must agree on a transaction validity. Examples of such algorithms are the Byzantine Paxos algorithm [105], the Practical byzantine fault tolerance algorithm [106] and variants that address the robustness such as Ardvark [107] and RBFT [108] or that address the performance problems of PBFT, such as Q/U [109], HQ [110], Zyzzyva [111] and ABsTRACTs [112]. An interesting byzantine fault tolerant distributed commit protocol is proposed in [113], where the authors enhance the classical 2-Phase Commit protocol by replicating the coordinator to successfully terminate when the coordinator failed and by building a quorum of coordinators to validate transactions and identify malicious participants.

The Proof Protocols are used by most of the public DLT systems [46,47] for supporting the consensus mechanisms in order to ensure the consistency of the ledger state across the network nodes. The Proof Protocols have defined two categories of nodes: Provers and Verifiers, where the Prover who may have unlimited resources needed to convince Verifier nodes with limited resources, about the truthfulness of a statement. As opposed to traditional Byzantine Agreement (BA) protocols, that use a quorum of participants to validate a transaction by voting, the Proof-of-Work algorithm validates a transaction (or a set of transactions) by solving a computationally intensive problem by a Prover that require a lot of physical resources and make infeasible for an attacker to cast an erroneous vote. The time needed by the prover to solve the computationally intensive problem gives the mining rate and directly influence the throughput (number of transactions) and the network scalability.

From the initial implementation of PoW in Bitcoin, where one block was generated every 10 minutes, PoW variations have been proposed aimed at improving the mining rate in order to obtain a higher throughput of transactions per second. The Greedy Heaviest Observed Subtree (GHOST) protocol [48] proposed by Ethereum increases the mining rate from 1 block per 10 minutes to 1 block per ~15 seconds. In order to avoid the potential problems that may arise due to delayed propagation of blocks, GHOST uses references to orphan blocks or Uncles (valid blocks that were not accepted in the main chain due to network delays) in order to increase the weight of the longest chain. In this sense, each new block can contain references to previous Uncles and for each of the referenced uncles, the miner will receive a small incentive, consequently, the miner of the uncle will also be rewarded when a new block makes a reference to it. This mechanism discourages the faulty miners to mine on forked chains and from perusing long-range attacks. Other variations of PoW have been considered in order to impose some restrictions on the hardware devices used for mining by encouraging the implementation of ASIC (Application Specific Integrated Circuits) resistant algorithms for hashing. This came as a result of the Bitcoin's early years when hardware companies started to profit from the popularity of blockchain solutions by developing ASICs in order to increase the hash rate of the computing nodes.

However, one such circuit may cost around 3000 dollars [49], which makes it unprofitable for a simple user to invest in such hardware and gives more power and control to large companies and to the manufacturer. In order to avoid this problem, the next generation of DLT solutions researched and applied new hash functions that are ASIC resistant. ASIC resistant algorithms try to shift their strategy from CPU intensive algorithms to memory intensive algorithms, called Memory hard puzzles. This came because the performance of processors has increased over time at an exponential rate, as opposed to the memory which has known a more linear increase. The purpose of these algorithms is to design a method that requires large amounts of data to be stored, that cannot be efficiently parallelized. Scrypt [50] is one of the first ASIC resistant algorithms and is currently widely used by many applications. However, Litecoin, which is one of the top platforms that use this algorithm set the memory size at 128 KB [51] thus making it possible to be stored at the CPU cache level. This restriction was applied since the Scrypt algorithm requires the same resources for solution verification as for the solution discovery and higher requirements would stress too much the regular non-mining nodes. Dagger Hashimoto [52] [53] on the other hand, is an algorithm that provides an easy verification solution, thus allowing the Prover's requirements in memory size to increase up to 1 GB RAM. Equihash is also a widely used hashing algorithm. However, the main disadvantage, as the authors themselves state [54], is that the algorithm is parallelizable, which is not a quality desired in ASIC resistant algorithm. Finally, the Cuckoo hash cycles [55], used in [56] and [57], is also considered a reasonable solution when talking about ASIC resistance. Other relevant variations of PoW algorithms aim at giving a purpose for all the energy and computational resources of the network. Since the network uses large computational resources whose only purpose is to prove and validate the next block of the blockchain, the concept of Proof of Useful Work is launched as an alternative to trying to use the computational power for a publicly beneficial domain. Such implementations aim to do research work (or Proof-of-Research). They gather the computational power across the network in order to provide solutions to some of the world's problems. CureCoin [58] is implementing such an algorithm called SigmaX that aim to perform protein unfolding in order to find a cure for different diseases.

Virtual Mining Protocols offer an alternative to the PoW by keeping a high cost for the Prover, but changing the resource consumed. If the cost of the Prover in PoW is the energy consumed, which would be lost if the Prover does not offer honest work to be validated and rewarded by the network, in the virtual mining Protocols the cost is a deposit of coins that are offered as insurance for their honest work. If up until now the node was chosen based on its result to the computationally intensive problem, now the node will be elected in a pseudo-random way, and the chance of winning will be proportional to the number of coins / stakes of the owner of the system. Thus, in Virtual Mining Protocols, the clients have the mining potential proportional to the percentage of the stake they hold.

Three virtual mining approaches have been identified across different solutions: Proof of Stake considers the age of the coin in the algorithm, thus requiring for some coins not to be spent for a period of time; Proof of Burn requires for a relevant amount of coins to be destroyed and a proof of the destroying transaction to be provided; Proof of Deposit requires for some coins to be put away for some time in a vault. However, all three algorithms have the same purpose that is, incentivizing the honest work of the miner by promising as a reward a sum of coins greater than the initial insurance. Proof of Activity [59] is a hybrid algorithm build upon Proof of Work and Proof of Stake found in Decred [60]. The algorithm starts as a simple Proof of Work algorithm until one correct hash is found; the block is then transmitted in the network, but it is not yet added to the blockchain. In order to become a valid block, it needs to be signed by N holders in the network. The PoW obtained hash is used to generate N numbers that correspond to N coins generated since the genesis of the blockchain. Each of these coins has one current stakeholder who will be required to sign the current block. The signature of all the N stakeholders is required in order to consider the block valid. In case that some of the stakeholders are not online and cannot sign, then the miners will continue their job in order to find a new hash and ask other stakeholders to sign the block. This approach makes attacks upon the network more difficult since it makes use of the advantages brought by both systems.

According to [61], the Casper version of Proof-of-Stake (PoS) is considered a suitable alternative for the permissioned systems, by considering only a fixed set of users as validators of blocks. Another flavor of Proof-of-Stake commonly used for permissioned systems is the Delegated Proof of Stake (DPoS). In DPoS, N witnesses are periodically selected by stakeholders of the system, such that enough decentralization to be ensured. Out of the N witnesses, each witness has its chance to propose the next block, and then be rewarded for its contribution. Proof of Authority, on the other hand, suggests that only trusted parties are entitled to provide commits to the system, which can be required where high-security properties need to be implemented, like in the case of private Enterprise solutions. In Quorum, RAFT algorithm is used, where a predetermined leader is creating a block that is sent to each node in the cluster [62]. In terms of finality, proof-protocols are known not to be final, however they offer probabilistic finality, since once many blocks are sealed over, the probability of a block's state to change is very low.

From existing Virtual Mining Protocols, the Proof-of-Stake has a good potential of becoming the most used consensus protocol in DLTs because it addresses fundamental problems of the PoW protocol such as computational waste and high-power demand [114]. Anyway, in case of PoS algorithm, since the nodes propose a new block by guaranteeing with their own stake it gives rise to the "nothing-at-stake" vulnerability. This means that when a fork appears in the context of a network partitioning, an attacker node can propose a block on either chain, hoping that at least one block will be accepted. The node guarantees each proposed block with its own stake, but due to network partitioning it is difficult for other nodes to observe and penalize this misbehavior. This situation can lead to other forks or to the fact that the attacker node receives rewards for proposing new blocks. In PoW algorithms, the "nothing-at-stake" vulnerability is avoided due to the fact that when proposing a new block, the node has to solve a computational puzzle that consumes electrical energy, and by proposing two blocks on two chains from a fork means that the node has to solve twice the problem, thus doubling its costs.

There are two categories of PoS mechanism: i) chain based PoS that mimics PoW by assigning pseudo randomly the right to generate new blocks to various nodes and ii) Byzantine Fault Tolerant PoS that is based on BFT research. They address the "nothing-at-stake" vulnerability in different ways. The chain based PoS are penalizing nodes when sending multiple blocks on competing chains (e.g. Slasher [115], [116] or Casper [61]). The BFT PoS mechanisms allow validators to vote on blocks by casting several messages, with two rules: finality condition (to determine when a hash is finalized) and slashing conditions (to determine when a validator misbehaved and must be excluded). A block is considered finalized once enough votes have been cast and all nodes from the DTL agree on adding it to the canonical history. This involves sending many messages in the network to make aware other nodes that a new block was proposed and running a version of Byzantine Agreement on the new block. Propagating many messages in the network impacts system scalability, thus methods to reduce the number of messages is needed. Two techniques are found in literature addressing this: i) quorum based voting – when a node is selected randomly as the prover and a subset of nodes are selected to be verifiers that run a Byzantine Agreement protocol (Algorand [114]); and ii) sharding-based approaches – where the blockchain is split into shards for inter-shard transactions and only transactions that involve nodes from two different shards need message propagation between shards (Casper [61], Elrond [117]). Algorand is based on a new and fast Byzantine Agreement Protocol used to generate a new block through a binary Byzantine Agreement (BA*) protocol that enhances the traditional BA protocol to work in rounds in a synchronous environment with at least 2/3 players being honest. Furthermore, a cryptographic sortition based on Random Verifiable Functions is used to select a subset of the users to be members of the BA* algorithm. A cryptographic function is used to select a new leader based on a previous block. The leader will be in charge to propose the new block. A set of verifiers is used to check the validity of the new proposed block. The choice of the leader is not predictable, thus making impossible for an attacker to alter the new block. Furthermore, leaders learn of their role without informing others only after proposing the new block, thus avoiding attacks. After a new block is proposed, the leader has no importance for the algorithm. However, the verifiers must agree on the new block, and they run the BA* algorithm in rounds, at each step players being replaced, thus avoiding cases when many verifiers are corrupt. Elrond is based on a sharding approach, splitting the blockchain and account state in several shards where parallel validation can occur using a consensus algorithm based on a secure PoS. The consensus algorithm follows a similar approach as Algorand with a prover and a set of validators chosen randomly within a shard and running a Byzantine Agreement algorithm to validate the proposed block. Finally, Hot Stuff [118] proposes a consensus algorithm using a leader-based Byzantine fault-tolerance protocol for partially synchronous distributed system models where a chosen leader drives the consensus decision at the rate of the maximum delay allowed by the network. Table 6 shows a comparison between main consensus algorithms considering the relevant features discussed above.

**Table 6.** Consensus Algorithm Comparison

| Features | Bitcoin Proof of Work | BFT based Proof of Stake | | | |
|---|---|---|---|---|---|
| | | Casper FFG and CBC | Algorand | Elrond | Hot Stuff |
| **Partition Resilient** | Yes, eventually consistent | Partition resilient and available with a tradeoff in consistency, leading to attacks such as Nothing-at-Stake | | | |
| **Network assumptions** | | Permission-less | | | Permissioned |
| | Synchronous network | Synchronous network | Synchronous network | Synchronous within shards Asynchronous cross-shard | Partial synchrony model |
| **Maximum faulty nodes in the system** | 49% | 33% | 33% | 33% | 33% |
| **Scalability (Reported Network Size)** | >10.000 full nodes | - | 50-500K nodes | 16 shards, total number of nodes N/A | 128 nodes |
| **Transaction throughput** | 2 TPS (1 block every 10 minutes) | 15 TPS | 250 TPS (1 block in less than 10 minutes) | >10.000 TPS | 50 TPS |
| **Transaction Finality** | More than 1 hour (6 blocks) | Yes | 1 block in less than 10 minutes | 1 block in less than 10 minutes | Yes - Proven |
| **Smart Contracts** | No | Yes | Yes, but not Turing complete | Yes - EVM compliant engine | No |
| **Sharding** | No | Yes | No | Yes | No |
| **Prover** | First Miner that solves the computational problem | Node from Dynamic Validator Set | Node chosen randomly using Verifiable Random Functions and last block hash | Block proposer from an eligible set committing stake | Leader chosen from the network, center of the star communication network. |
| **Validator** | Any other full node can check the transactions from the newly proposed block | Dynamic Validator Set selected according to stake. Nodes can join and leave the set dynamically. | Validator set chosen randomly using Verifiable Random Functions and last block hash. | Other nodes from eligible set | Nodes from the validator set |
| **How nodes agree on new block** | Each full node can validate the block | Message passing protocol Byzantine Agreement | Byzantine Agreement run by validators | Modified Practical Byzantine Fault Tolerance | Modified Practical Byzantine Fault Tolerance run in three phases. |
| **Generation of new tokens** | Yes, every miner is rewarded several bitcoins. | Yes | No need for incentives for validators | Yes (ERD) | - |
| **Existence of Forks** | Yes, miners will propose blocks only on longest chains from forks. | Yes | No – Only Adversaries can create forks, but it has a small probability | Yes – Only within shards; Forks of maximum 2 blocks long. | - |
| **Protocol Finality** | Yes, mining process takes a finite time | Ongoing research | Yes - Byzantine Agreement | Modified Practical Byzantine Fault Tolerance | Yes - Proven |

## 3. Scaling Tier

The DLT scalability limitations are mainly influenced by the restrictions imposed by the consensus algorithms on the Protocol and Network Tier solution. These limitations are imposed either by maximum block size (e.g. 1 MB Bitcoin) or by a cost constraint (e.g. gas consumption and gas price in Ethereum). Combining these constraints with the strict periodicity of the block generation (e.g. 10 minutes for Bitcoin, 15 seconds for Ethereum) results in a system that can process a limited amount of transactions impacting both the storage and the processing capabilities (due to the gas consumption costs in case of Ethereum smart contract execution). Bitcoin reportedly can allow 7 transactions/second on average [68], while Ethereum registers 13 standard transactions/second or 7 transactions/second in case smart contract execution is involved [69]. Even private deployments reach certain limitations, Hyperledger advertising 100 000 transactions/second, although reports show a stricter limitation of 700 transactions/second [70]. In order to achieve higher scalability, one option would be to curtail the strength of Protocol and Network Tier DLTs by compromising the security, the immutability or the consensus of the system. Since this would not be an acceptable solution, the scalability challenges opened new research perspectives for the DLT usages. In this sense existing systems or concepts (databases, file systems, etc.) have been reconsidered and integrated with Protocol and Network Tier solutions, generating the Scaling Tier solutions that can combine the benefits of more traditional solutions with the ones of the DLT systems.

## 3.1. Storage Size

Storage scalability is a requirement of many use cases nowadays. With the increase storage capabilities of the traditional systems, much of the paper documentation have been digitalized, leading to a world of digitalized records in domains like healthcare, intellectual properties, real estate, legislative contracts, etc. Furthermore, the media and social network use cases are more and more flexible, providing increased storage options for users to store their files (documents, photos, videos, etc.). DLTs caught the interest of these domains, aiming to maximize their potential, by ensuring immutability (legislative and real estate), provenance (intellectual properties), security (healthcare), etc. However, the greatest challenge of integrating DLT solutions with these domains is limited storage capabilities.

In order to achieve storage scalability several solutions have been proposed, either by providing a hybrid architecture of Protocol and Network Tier solutions (e.g. Sharding) or by integrating well-known systems (e.g. file systems) with existing DLTs. All the Scaling Tier solutions aim at storing all the data off-chain and keep only a digital fingerprint of the data on the Protocol and Network Tier. While the data kept on the Protocol and Network Tier benefits of all the advantages the system provides (consensus, immutability, security, etc.) it is considered a source of truth in the validation of data that is stored on the Scaling Tier.

Sharding is a solution implemented in order to improve the storage scalability on-chain. Different nodes are assigned to process and store only a corresponding sub-category of transactions. A simple sharding technique is to split the network in shards corresponding to the transaction's prefix: 0x01 shard, 0x02 shard, etc. In the sharding mechanism proposed by Ethereum Sharding [71], the system defines objects at three different levels: level 0 – transactions; level 1 – collations; level 2 – blocks. The collations are the data structures responsible for package transactions that belong to a shard. The collations are created and sealed by Collators that are nodes in the network registered on the main chain in the Validator Manager Contract. The Collator deposits a sum of coins on the main chain based on which they will be chosen in a Proof of Stake manner to validate the next collation. The header of the proposed collation will then be verified on-chain and added in the next block on the main chain. Cross-Sharding communication is also possible by providing Merkle-Proofs of existing transactions from the main chain. A similar approach to the Ethereum Sharding is investigated by Distributed Technology Research Foundation [72], Elrond Network [74], Hyperledger [16], Elastico [74], Omniledger [75] and Rapidchain [76]. The Scaling Tier solutions that use file systems as storage mechanism allow large files to be stored by fragmenting, encrypting and sharing chunks of the original file between the nodes of the Scaling Tier network, while the hash of the original file is stored in the Protocol and Network Tier. The farmers (nodes storing the data) need to respond to periodic checks regarding the integrity of the stored data, and a reward scheme is implemented for their services. The storage mechanism and integration with the Protocol and Network Tier is presented in Figure 4. There are several successful implementations of Scaling Tier distributed file systems among which, worth mentioning are: Storj [77], IPFS [78], Filecoin [79], Decent [80], Sia [81], MadeSAFe [82], Swarm [83] and Arweave [84].

IPFS (InterPlanetary File System) is one of the most used Scaling Tier solutions for file storage. When a user publishes a large file using its own IPFS node, the node will first fragment the file in smaller chunks, the hash of each chunk becoming a node in a Merkle DAG, whose root is the hash of the initial file, thus making use of hash pointers in order to ensure tamper-evidence. For security reasons, the chunks stored are of standardized sizes, so that an attacker cannot extract any useful information by analyzing the size of a chunk. The owner of the data is responsible to hold the private key used to encrypt the chunks of data that are scattered across the network. This makes the system highly secured since even the data is stored across multiple nodes, the mechanisms make it impossible for anyone holding the data to use it since it is encrypted and fragmented. Also, it ensures security through encryption and no downtime since the file is shared across multiple users. The system offers the possibility to transfer data, check the availability and the integrity of the stored data, retrieve the data and pay for the service provided. While the payment scheme is not integrated into IPFS, similar implementations have emerged: Storj and Filecoin, proposing to reward and motivate the decentralized nodes to act honestly regarding their storage services. Ethereum Swarm, similarly to IPFS, is a peer-to-peer system that aims to store data in a decentralized way and relies on immutable content¬-addressable data. While IPFS needs Filecoin to validate storage proofs, Ethereum Swarm proofs are validated at the contract level and rely on incentive schemes based on the native coin, Ether.

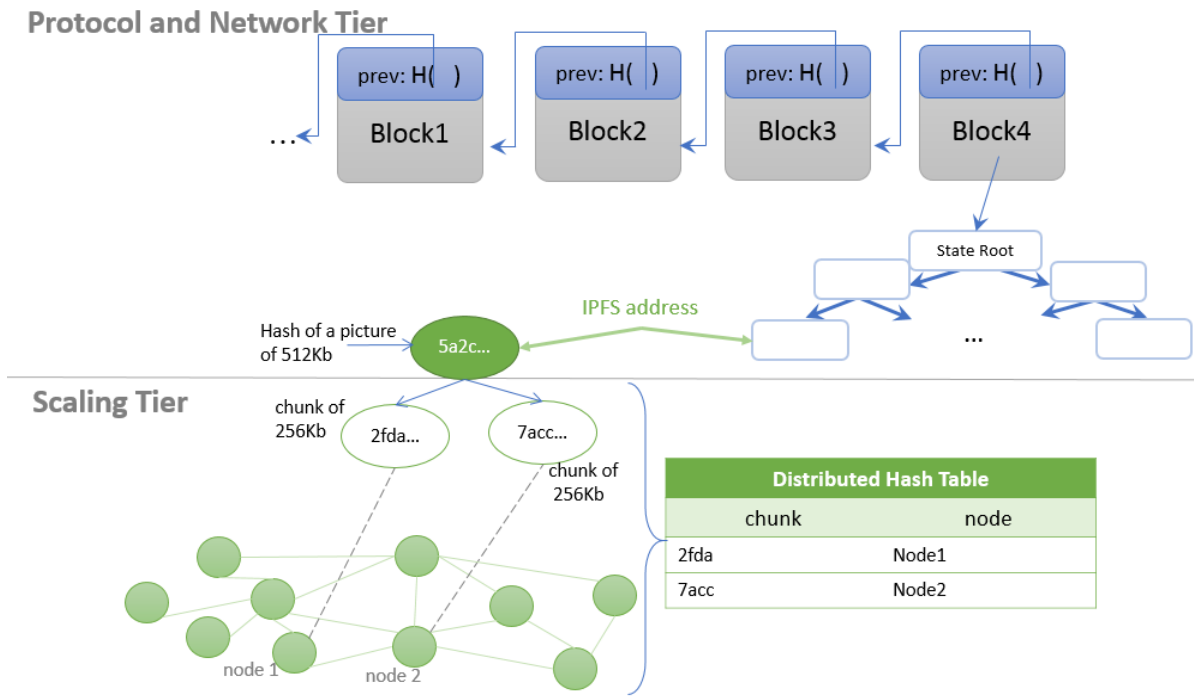

**Figure 4.** Scaling Tier - File System Storage Mechanism

In Table 7 a comparison between the identified storage scalability solutions is presented. Sharding presents a promising alternative to the classic DLTs, by providing increased storage capabilities.

**Table 7.** Scaling tier storage solutions

| Features | Protocol & Network Tier | Scaling Tier | |
| --- | --- | --- | --- |
| | Fully Replicated | Sharding | File Systems |
| Immutability | Yes | Yes | No |
| Trusted Parties | None | None | Peer nodes |
| Byzantine Tolerant | Yes | A tradeoff with the no. of shards | No |
| Storage Scalability | Low | Medium | High |
| Cost | High | Medium | medium |

If the classic DLT has the storage capacity bounded by the block size and block generation periodicity, in sharding this storage capacity is multiplied by the number of shards, by having the block sealing process parallelized with each shard. However, by increasing the number of shards, fewer nodes get to be assigned per-shard for validation. This can easily make the network susceptible to attacks since by attacking one shard the entire system can be compromised. A Scaling Tier file systems solution however, even if it provides great scalability in terms of storage, it also requires a degree of trust between the storing nodes. Since IPFS is not fault-tolerant on its own, the DLT storing the hash ensure only tamper-evidence in the system but does not make the system tamper-proof. Furthermore, each time an update is applied to one of the files, the hash pointer changes as well, requiring a transaction updating the entry on-chain as well. While this is desired in order to keep a tamper-evident system, a high frequency of updates will also lead to higher costs.

### 3.2. Transactions Throughput

Delivering a solution to the low transaction throughput of the DLT systems may lead to unlocking the disruptive blockchain technology potential over areas where micro-payments are exploited (e.g. Energy Sector, Media Services, etc.). Micro-payments are online transactions involving small amounts of money. These small amounts of money are often used in exchange for different goods or services, and most of the times require many transactions over a period. The main problems that arise in DLTs systems by integrating micro-payments are the high fees accumulated as result of the mining fees paid for each small value transaction and the congestions problems at the level of the Protocol and Network Tier due to a large number of transactions and the small size of the block. Sharding can be considered a suitable solution for increasing the transaction throughput as well as the storage. Both improvements come because of introducing clusters of nodes responsible for specific categories of transactions. By delegating the transactions to a different category of nodes in the system and parallelizing the validation and sealing of these transactions, a system with higher transaction throughput is provided.

Sidechains [85, 86] are proposed as alternative for increasing the scalability of the DLT solutions. A side chain is processing transactions in parallel with the main chain while the transactions of the sidechains are always rooted in a locking transaction in the main chain. Once proof of a locking transaction is made on the side chain, the actors may start using the assets by transacting on the side chain. In order to return to the main chain, a proof of the latest state from the sidechain must be made in order to unlock the coins on the main chain. There are different reasons for connecting to side chains: testing new functionalities, extending the main chain functionalities (e.g. RootStock [87] enabling smart contract execution), or moving business-specific implementation to another less expensive chain (e.g. Plasma as a tree of sidechains [45]). Either the case, by moving part of the transactions from the main chain to side chains can have as effect an improvement in the transaction throughput, since part of the transactional load is moved for a period outside the main chain.

In terms of transaction scalability, the payment channels, mainly implemented in Lightning Network solutions [88] are one of the best choices for improving transaction throughput. The payments channels aim to combine all the small payments into one large payment at the end of a service period. The Lightning Network provides a point-to-point network that runs on top of the blockchain network as depicted in Figure 5 and relies on hash time locks and cryptographic secrets to ensure the reliability of off-chain payments. The nodes in the Lightning Network exchange cryptographic promises (signed transactions) with each other that represent the payments exchanged between them offline and that can, at any point be deployed on the main chain in order to securely redeem the associated coins.

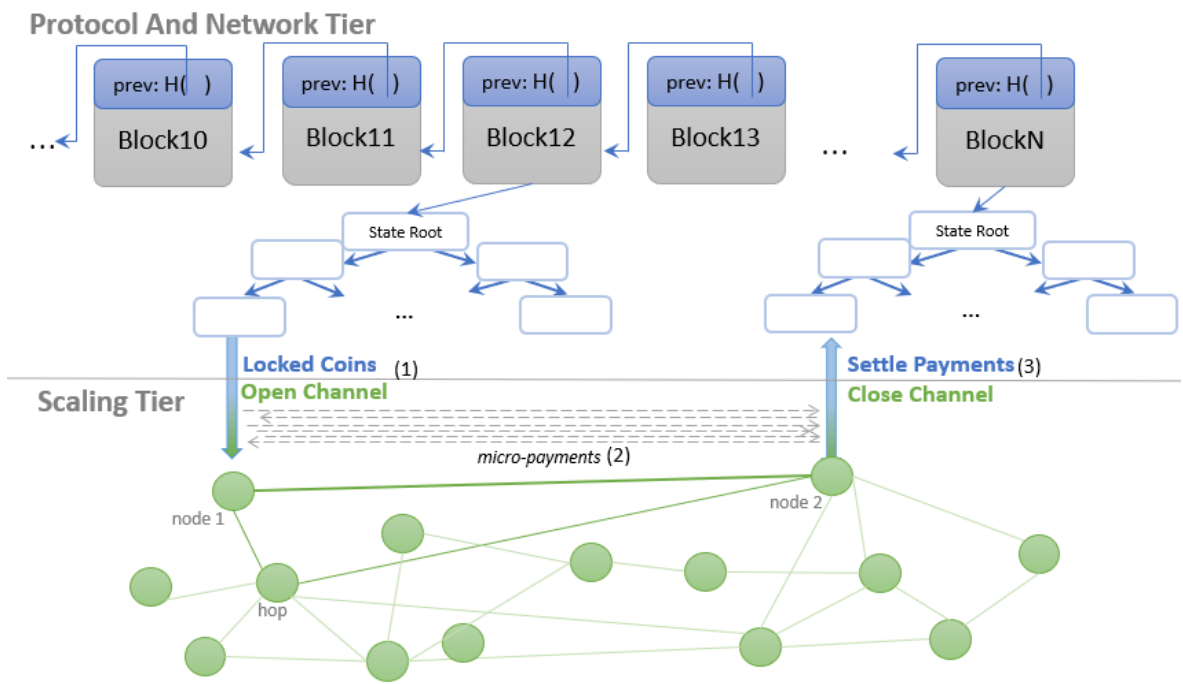

**Figure 5.** Scaling Tier - Payment Channels Mechanism

Whenever two parties aim at exchanging many micro-transactions, the first step is to open a channel between them, represented on the DLT as an opening transaction that locks the maximum amount of money that the two parties could transact off-chain. This opening transaction is signed by both parties, thus unlocking it will require both parties to agree on the transaction spending the locked amount. After the initial opening transaction, the parties will continue to exchange messages off-chain, namely, commitment transactions, which are designed as fail-safe mechanisms against cheating off-chain. Whenever a transfer occurs, determining a change in balances, each party creates one commitment transaction specifying the updated balances, signs it and sends it offline to the other party. Upon receiving the commitment any party can successively sign it and publish it on-chain to redeem the coins or can wait and make other off-chain transfers and return only with a future commitment transaction on-chain. In order to prevent any party to return to the chain with a deprecated and more favorable commitment transaction, a secret-locking and time-locking mechanisms are incorporated in the commitment transaction that provides a fail-safe mechanism such that any party acting fraudulently risks losing all the money in favor of the other channel party.

A lightning network is a peer-to-peer network that usually exists in parallel with the main DLT network. It provides a mesh of the previously described payment channels that allows any two users from the main network to securely exchange offline transactions by sending them through a payment channel opened directly between the transacting parties, or by making use of a route (or path) existing in the graph formed by nodes and opened channels (edges). The benefit of transferring over a route is the reduction of the cost associated with the opening transactions required on-chain each time a channel is required with a specific party. For sending the transactions through a path that requires intermediary parties (hop nodes) to route the messages, additional security mechanisms are required to ensure the correct delivery of the transacted assets. For successfully issuing a routed transfer, a commitment transaction is exchanged between every two parties involved in the path channels. A Hash Time-Locked Contracts have been defined over the previously presented mechanism, in order to prevent the intermediary to unlock the routed commitments before the receiver can confirm the payment. Upon confirmation, a proof is issued by the recipient and sent to the intermediary in order to unlock the hash-locked commitment. If the proof is not provided during an established period (time lock) then the intermediary will no longer be able to claim the payment. The mechanisms of the Lightning Networks were firstly defined for Bitcoin in [88], and future implementations have followed for another networks Raiden for Ethereum [89], Bolt for Zcash [91].

In Table 8 the three systems offering transaction scalability are compared. While sharding can be considered a suitable solution, an additional issue arises when increasing the number of shards and of transactions. When a transaction is sealed by a shard, it may reference a transaction that is not in the log of collations of that specific shard. As a result, each time such a transaction needs to be validated, a cross-shard communication is required in order to issue Merkle Proofs of the referenced transaction. Consequently, increasing the number of shards and of transactions will also introduce a communication overhead that may impact the scalability of the solution. More exact evaluations of the overhead introduced will be possible only after these currently researched solutions will offer a full specification and deployment on the public networks.

**Table 8.** Transaction scalability solutions

|  | Protocol and Network Tier System |  | Scaling Tier System |  |
|---|---|---|---|---|
|  | Fully Replicated | Sharding | Sidechains/Plasma | Lightning Network |
| Trusted Parties | None | None | Depending on the implementation | None |
| Transaction Scalability | Low | Medium | Medium | High |
| Cost | High | Medium | Medium | Low |

The sidechain implementation can be considered to offer an alternative to the sharding shortcomings, by fully relying on the information stored on the side chain and integrating with the main chain only when locking and unlocking the coins. However, from a scalability perspective, while this solution may offer some improvements to the underlying DLT, by taking over some of the transitioning load, the transaction throughput is still limited since the side chain is most of the time a DLT with the same constraints as any Protocol and Network Tier solutions. Additionally, some issues regarding the overall security of the systems need to be considered, since upon returning on the main chain, the transactions validated by the side chains are considered to be valid even if the validators' network of the side chain differs from the ones of the main chain.

The Lightning Network offers the best solutions by being implemented without relying on any third parties and by providing higher transaction scalability compared to other solutions, promising millions of transactions per second. However, the problem studied is that once the network reaches many users it gets more difficult to find a viable route between two parties. The most desired topology of the lightning network would be a completely decentralized network, where each node has connections with other nodes, together forming a mesh of channels that are part of a connected graph. However, in order to be able to route money from one point to another, all the nodes from the path need to have at least the same amount of money as the one requested by the initiator. Taking advantage of this issue motivates some actors of the system to open channels with a larger number of parties and fuel the channels with enough money to be able to route and connect different parts of the network, acting like large routing hubs in exchange for small fees.

### 3.3. Processing Capability

In DLT systems that allows complex functionality and computations through smart contracts, there is an important restriction, that although was introduced for security reasons, it can be considered a limitation for more complex operations. The concept of gas was introduced in Solidity in order to measure the amount of computational effort. When running smart contracts, each executed operation and processed byte of data is paid for (gas multiplied by the gas price). When deploying transactions on-chain, that are triggering functions from the smart contracts, the issuer must also have enough coins for sustaining the operations thus rewarding the miner for using its computational resources. At the same time, this mechanism prevents an attacker from running extremely long tasks or infinite loops, since the attacker would need to provide enough reward to incentivize the miner to execute each operation. When the reward provided runs out, the computation stops, and the transaction is dropped, thus avoiding situations where the nodes become unavailable due to attacks or complex computations. By integrating the Protocol and Network Tier system with external services, this shortcoming can be solved. The nowadays solutions for addressing complex computational problems in DLT are Oracles [91] and Proof of Computation mechanisms [92].

The Oracles are mechanisms that provide a secure connection between the chain and the outside world becoming a trusted third-party entity, or a network of entities, for the Protocol and Network Tier. The Oracles can be used to offer results from different URLs, [93], IPFS, or units responsible for running more complex algorithms. The problem with interacting with the outside world, directly from the chain, is that the response must be the same across any number of requests issued by the nodes during mining. This proves to be almost impossible when accessing dynamic changing data regarding weather, stock prices, etc. One problem that can appear the Oracle-based system is data tampering or man in the middle attack. In this sense, the Oracles are responsible to ensure the authenticity of data through authenticity proofs. One problem that persists is the centralized nature of the Oracles. The mechanism is presented in Figure 6, where an event containing details about the request is issued and intercepted by the Oracle. The necessary information is retrieved from external services and published back on the chain through a callback transaction. A similar concept is proposed in [94].

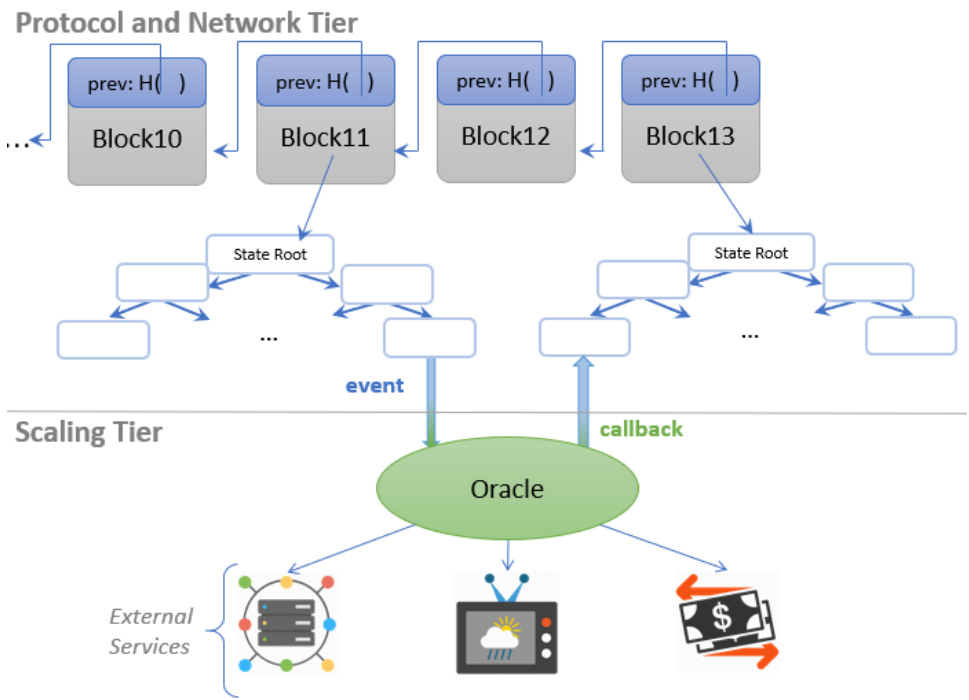

**Figure 6.** Scaling Tier - Oracle mechanism

Other implementations aim to outsource computing-intensive problems to off-chain nodes by implementing a proof-of-computation mechanism as depicted in Figure 7. TrueBit [92] is such a system that relies on Ethereum Smart Contracts and gives the possibility of peers to request solutions for complex tasks. A Solver that has enough computational resources will run the tasks outside the chain and submit its results.

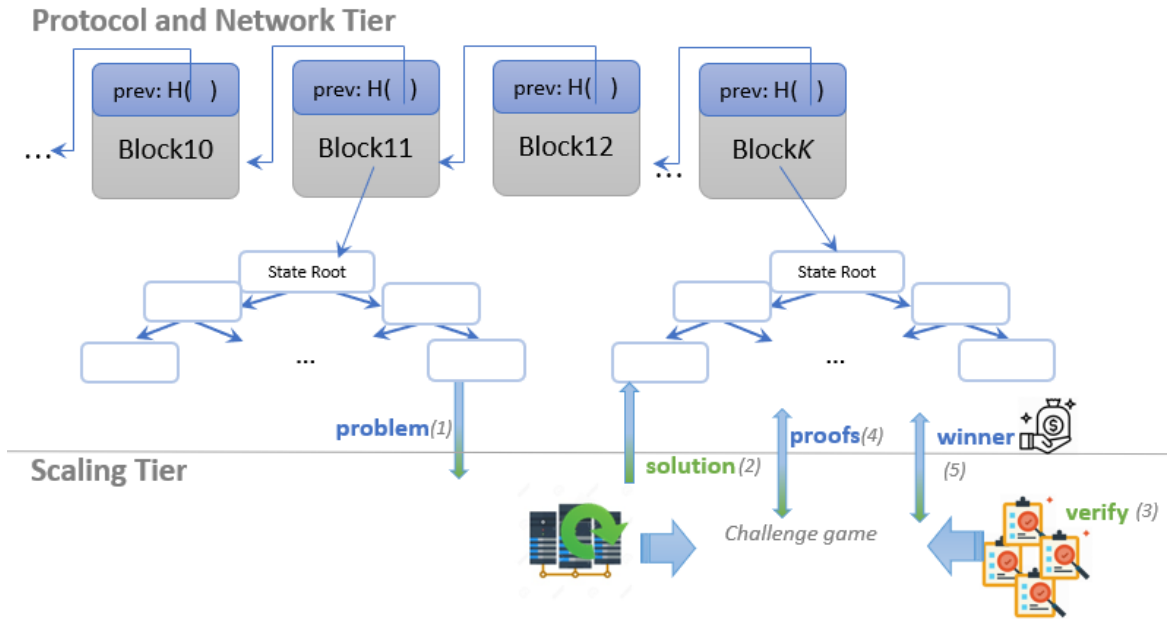

**Figure 7.** Scaling Tier - Proof of computation mechanism

Several Verifiers peers can evaluate the results, and if a disagreement occurs, a Challenger can contradict the result published, by starting a Challenge Game. Several rounds of proofs are registered on the chain in order to check whether the computation was done correctly. In the end, the winning part will receive a reward for its cooperation, while the part proven to be wrong will be charged for its actions. Another set of actors are de Judges that given the proof of the solution can easily verify the correctness of the game. However, a big drawback of the system is that it is limited to running tasks written in WASM [95]. Enigma [96] proposes a similar concept, of outsourcing the computation but with an added layer of privacy over the data and computation performed. By leveraging on secure multi-party computation, the proposed solution distributes the data across several nodes for computation. As a result, no central node will have access to the entire problem or solution, but to a seemingly unintelligible part of it. The proposed solution is not yet released in production, but it is currently tested on the Ethereum testnets [97].

Compared to the Oracles the Proof of Computation is a better choice since it is implemented without relying on any trusted party, and at the same time provides validation of the result implemented directly on-chain. The proposed solutions show great potential in the future, that may disrupt many use cases which require security and correctness to be applied on more complex computation than the ones that can be handled on-chain.

## 4. Federated Ledgers

By launching various domain specific DLT systems for public use, each addressing different requirements, one problem that arises is the need of a protocol or interoperability mechanism that can allow the transition from isolated DLTs to networks of DLT. Such a network could further unlock the potential of the DLT systems by allowing one business to leverage on the potential of different DLTs for different categories of operations.

Several solutions have been investigated whenever a transfer of coins was needed from a parent DLT to a secondary DLT. This type of Ledger-to-Ledger communication is mostly ensured through two-way peg systems [86]. The transfer is achieved by temporary locking some coins on the parent chain and then unlocking the same amount of coins on the secondary chain. The transacting process is executed on the secondary chain until the user decides to return to the main chain. This is possible, by repeating the same process on the secondary chain, thus locking the coins on the secondary chain and then releasing them on the main chain.

One proposed solution to implement this swap mechanism is by having a central exchange [87]. This requires the central exchange to have access to both chains, thus once a user sends a request from the main chain specifying the amount of money and the address that should receive the coins in the secondary chain, the exchange can then easily send the same amount of coins at the specified address on the secondary chain. The drawback of this approach is that the central exchange needs to be a trusted entity; otherwise, this exchange could easily steal all the money from the two chains. As an improvement, a MultiSignature Federation can be implemented. In this case, any transaction must be approved by N out of M participants, instead of relying on only one entity. It still relies on a middleman, but the risks are reduced.

Another solution used is to entangle to chains relying on the secondary chain to monitor the main chain and includes all the block headers in the current's chain blocks. As a result, blocks from the secondary chain will have two parents, the previous block from the secondary chain and the last mined block on the main chain. However, a big disadvantage of this approach is that the primary chain needs to create blocks at a lower rate than the secondary chain. An entangled model is used by BTC-relay that connects the Bitcoin chain with the Ethereum chain. After every 10 minutes, proof of the last mined block in Bitcoin is provided in the Ethereum smart contract. Each time a user wants to prove the validity of a Bitcoin transaction on the Ethereum chain, it only needs to provide the Merkle Path of the transaction and the block it is contained in (a Simplified Payment Verification (SPV) proof).

The sidechain approach aims to eliminate the middleman completely. However, the protocol of the chains must implement a new way of unlocking coins, that is proof of a locked transaction on the other chain. The transacting party will submit on the sidechain an SPV proof of the transaction deployed on the main chain. The SPV proof will firstly contain the Merkle path containing the proof that his transaction was submitted and mined, and furthermore, it will contain the hashes of all the blocks that followed. Upon receiving the proofs, the sidechain will enter a reorganization period. This period aims to provide the necessary time to avoid any possibility of double spending problem. In this period, in case of a fork in the main chain network, anyone can offer a new SPV proof that contains the same block as the one provided initially, but without the transaction in it. If the proof provides a longer list of block hashes that confirm the missing transaction block, it is concluded that a double-spending attack was committed on the main chain and the original transaction is ignored in the sidechain as well. Upon withdrawal on the main chain, the same process occurs. The main chain lockbox is defined as a new type of transaction that can only be unlocked using SPVs of locked transactions from the other chain. The system still has a notable drawback, since the main chain needs to rely on the integrity of the sidechain. If the miners of the sidechain are not honest and coins are stolen, the main chain has no way to prove the honesty of the requests if a valid SPV is provided. Furthermore, if the amount of coins is split on the sidechain, in order to retrieve the coins on the mainchain all the owners from the sidechain must provide the SPV, thus no partial consumption of the coins on the main chain is permitted.

Drivechain [98] is an improvement of the sidechain. The protocol is similar to the sidechain, until the step requiring withdrawal from the sidechain back to the main chain. At this point, the SPV proofs are not used directly to withdraw the coins on the chain. During the reorganization period, several withdraw proofs are joined together aiming to consume a given amount of coins from the main chain. The withdraw transaction id, (not the actual transaction) will then be mined in the next block on the main chain's coinbase transaction. The mined ID will be interpreted as the intent of spending the locked money but will give time (1008 blocks) for all the users from the sidechain to validate the transaction and give chance to the miners to vote whether the actual transaction should be mined on the mainchain. In this way, the miners of the sidechain are prevented to commit illegal transactions, since any commit to the main chain will first be evaluated by the corresponding actors and majority of miners. A hybrid-model Is proposed by Rootstock [87], where a combination of sidechain, Drivenchain and multi-signature federation is implemented. It is a hybrid implementation that uses sidechain functionality for passing from the main chain to the secondary chain. But when returning, a Drivechain approach is used, where not only miners have the right to vote but also specially delegated notaries that vouch for the integrity of the transactions.

The inter-ledger communication domain is of much interest for Ethereum as well since with the rapid development of various businesses over the public chain, they aim to offer increased scalability of the main chain by offering alternative chains per application. One of the most recent applications, CryptoKitties [7], is such an example of required scalability. The actual application is completely independent of the state and data stored on the Ethereum chain, thus there is no reason for it not to function independently. However, the economic value of the application tokens need to be maintained, thus the system should allow purchasing the tokens on the main chain, paying with actual ethers for acquiring decentralized applications specific tokens, and then moving the tokens on a separate chain in order to use them in the actual application.

Plasma [45] is an alternative proposal of the sidechains on Ethereum. The current project in development relies on a UTXO based sidechain. An equivalent transaction is generated on the sidechain (plasma chain) from nothing, giving the corresponding coins to the plasma chain user. One of the first differences is the validation mechanisms based on Fraud Proofs by periodically checking the main chain. If a user wants to spend its coins on the main chain in a fraudulent manner, its action can be proved to be malicious by submitting a proof of a spending transaction that was previously registered on the plasma chain. This would prevent the main chain transaction from being validated. Since a user can issue a spending transaction directly from the main chain this offers a great advantage over the classical sidechain approach. That is, in case the plasma chain is compromised, the users can still issue their withdrawals. Another difference is that the Plasma model is designed to allow for nested chains, thus creating a tree-structured system of chains that requires each user to monitor only the chains that can affect one's transactions.

Interledger [99] aims to develop a protocol for allowing communication across different ledger by using payment channels. The mechanism is called "AtomicSwap" and uses routes existing in the Lightning Network. Considering that there exists a hop that has channels opened on two networks, e.g. Bitcoin and Ethereum, upon a transfer the hop can agree to update its Bitcoin balance from the Receiver channel, in return for Ether on the Sender channel.

In Table 9 the main approaches regarding inter-ledger communication are presented, comparing the main feature of each system: deposit and withdraw mechanisms, trusted parties and protocol independence. While hybrid models and Plasma may offer reliable solutions in terms of ledger-to-ledger interaction, the Lightning Network can be considered a viable alternative that can at the same time (being a Scaling Tier solution) allow very high transfer rates.

**Table 9.** Ledger to Ledger interoperability approaches

|  | Implementation Level | Deposit Mechanism | Withdraw Mechanism | Trusted Entities | Chain Independence |
|---|---|---|---|---|---|
| Central Exchange | Escrow | TX to a central authority | TX to a central authority | Central Authority | Yes |
| MultiSig Federation | Escrow | TX to a multi-signature federation | TX to a multi-signature federation | N Delegates | Yes |
| Entangled Chain | Protocol Layer | SPV Proof from the main chain | - | Dependent on the withdraw | Mining rate restriction |
| Sidechains | Network & Protocol Layer | SPV Proof from the main chain + proof of block validity | SPV Proof from the sidechain + proof of block validity | Sidechain miners | Yes |
| Drivechain | Network & Protocol Layer | SPV Proof from the main chain + proof of block validity | SPV Proof from the sidechain + proof of block validity + miners votes | Sidechain miners | Yes |
| Hybrid Models | Network & Protocol Layer | SPV Proof from the main chain + proof of block validity | SPV Proof from the sidechain + proof of block validity + miners votes + multi-signature notaries | Sidechain miners + Notaries | Yes |
| Plasma | Network & Protocol Layer | Proof for TX on the main chain | Direct withdraw + Fault Proofs | Central Authority, N Delegates | Yes |
| Lightning Network, Interledger | Scaling Tier Solution | Atomic Swap | Atomic Swap | None | Yes |

## 5. Discussion

As presented in the previous sections the blockchain technologies seen in the last years an unprecedented development bringing several benefits concerning a traditional based implementation. It eliminates the need for a mediator, using smart contracts, which enforce the rules on-chain, each side being aware of the consequences of their actions. The blockchain data structures allow for easy traceability of the assets and state updates that happen in the chain. Since the records are public and replicated this also provides great transparency so that companies and actions can easily be verified. Even if all the transactions and all the actions are public, the blockchain platform provides high security through consensus, public-key cryptography, and tamper-proof recording.

Nevertheless, a lot of nowadays applications and systems are rather centralized. Thus, efforts are being committed to investigating how the blockchain can be applied, integrated and used for decentralizing traditional systems such as medical systems, the provenance of products, Internet of Things, electricity grids, copyright management systems, autonomous vehicles, etc. In the rest of this section, we present a guideline or decentralizing existing systems by implementing new decentralized applications discussing relevant steps and technologies chooses. Figure 8 presents the proposed steps for decentralizing system design and business model using DLT and for implementing and launching it as decentralized application.

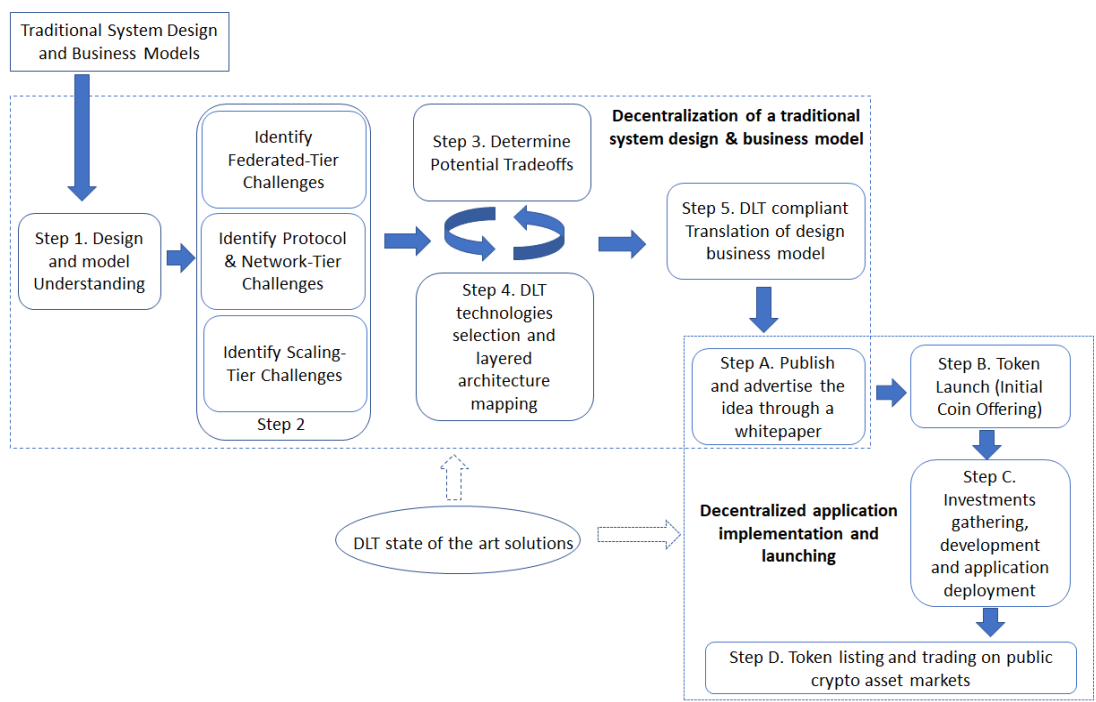

**Figure 8.** Decentralized application design and implementation steps

## 5.1. Decentralization of an application design

The main steps required for decentralizing an existing system by implementing decentralized applications are detailed below discussing their role and specific technological requirements (see Figure 8).

*Step 1.* Understanding the system centralized architecture and business model. To offer a DLT solution to an existing centralized model, an exhaustive understanding of the business is required. Having a good understanding of the business model and rules that govern the domain as well as the architecture used in the centralized implementation, are prerequisites for starting the sequence of steps required to achieve a consistent and well documented decentralized solution. At the same time, one needs to identify the main functional and nonfunctional requirements that the system needs to address for ensuring the completeness of the solution. Most of the time, a complete and sound list of requirements can be achieved by holding several interview sessions with the future stakeholders of the system.

*Step 2.* Identify potential challenges for PN-Tier as well as S-Tier. Once the functional and non-functional requirements are clearly defined, the challenges that can arise once mapping the model on a decentralized solution can already be identified. We classify the challenges based on the tiers identified at the level of architecture. Protocol & Network-Tier challenges: What is the targeted network level, public or private? Does the system need to hide the transacted information? Is selective disclosure a requirement of the business? etc. Scaling-Tier challenges: Are the scalability requirements higher than what a PN-Tier solution can offer? Are the scalability requirements targeting storage, throughput or computation? Federated Tier: What other DLT solutions are the system required to interact with?

*Step 3.* Determine potential tradeoffs. Depending on the set of challenges identified, there might be situations where not all the requirements can be successfully ensured by the DLT solutions. In such situations, a tradeoff needs to be made, since by moving the businesses from a centralized system, most of the time the scalability of the system may suffer in exchange for the reliability, immutability and consensus properties ensured by a DLT system. Furthermore, by outsourcing components of the system to the S-Tier, another tradeoff regarding the actual decentralization must be done, since most of the S-Tier solution requires a trusted party to oversee the outsourced component.

*Step 4.* DLT technologies selection and layered architecture mapping. Being a continuously researched domain, the DLT systems are very dynamic and a lot of research is done by both companies and academia to overcome the identified shortcomings and tradeoffs. In this sense, it is important that whenever a technological solution is considered for a model decentralization, thorough research needs to be done to have a clear picture of all potential alternatives to the problems identified. A DLT architecture of the system should be proposed, considering all the three tiers, and by selecting the most suitable solution for each architectural component.

*Step 5.* Present a DLT compliant translation of the design and business model. The re-implementation of the business model in a decentralized system uses a different paradigm as the one used in the classical centralized or web-based solutions. The key features of DLT programming, worth considering at this step are:

- The smart contracts can keep track of the data stored as well as act as a financial escrow for the parties interacting with it, by also having its own financial balance.
- Functions can be implemented to be used internally by the smart contract, or to be exposed as an API to the external modules. Each function exposed can receive two types of inputs: actual data used perform logic and update the state and payment values that can be required by the contract in exchange for performing different operations
- Most of the time the smart contracts are used as state machines, where the state is updated according to the latest input received. In order to interrogate a DLT for historical data, event-based information is considered. The events are structures emitted by the contracts during processing and stored in a dedicated Merkle Tree in the block.
- Each instruction executed by a smart contract has a processing cost that is paid by the transaction issuer. The payment is directed to the miner, for using its computational resources. Consequently, an appropriate data structure must always be used in order to avoid high costs. Furthermore, the processing rules should be implemented, if possible, in the same contract where the data is stored, in order to avoid the cost of referencing other contracts.

- When deploying a contract, the storage of the code is also paid by the issuer. For multiple deployments of the same contract, a good approach is to use libraries containing static code that is deployed only once and then linked to each of the deployed contract instances.

From this point forward, a classical pipeline of implementation, verification, and maintenance can be applied in order to validate the final application.

**5.2. Decentralized application implementation and launching**

Upon successfully implementation of above presented steps a decentralized design of the system and a DLT compliant translation of the business model is obtained. From this point forward one needs to proceed with the evaluation of the existing DLT implementation platforms, since some decentralized application may be compatible with existing systems, thus beneficiating from the already built network of miners. Two cases may emerge: an existing platform matches the decentralized application requirements, or a new custom chain needs to be implemented. In the first case Ethereum is such a platform, that allows the development of decentralized applications using Solidity and by deploying business-specific smart contracts on the public chain. Each new business idea that ends in being deployed on chain, usually is implemented around the business token or an asset representation. Once the business is launched, the higher the people's interest, the higher the price of the underlying token. Currently, there are 2667 registered decentralized applications only on the public Ethereum chain and 199 324 tokens registered [100]. In the second case a custom chain is built and use it as the base PN-Tier for the decentralized application implementation. Most of the times, this situation arises when the current frameworks' specifications are not compliant with the application requirements, thus a new DLT core platform must be developed and a new network of nodes must be built.

There are several steps to go through from launching such a decentralized application, from advertising, crowdfunding to the actual development and deployment (see Figure 8).

*Step A*. Publish and advertise the idea through a whitepaper. One principle of decentralized application whitepaper is to provide complete transparency to build trust with the investors. Companies are encouraged to provide an honest plan and roadmap from both a technical point of view (explaining in details the technical approaches and the feasibility of the solution) , and from an economic point of view (providing the investment plans, the shares among the company partners, the economic sustainability of the solution and the revenue plans).

*Step B*. Token Launch (Initial Coin Offering – ICO). A fixed number of tokens should be released to attract investors and raise money for the development phase. The token distribution plan specifies the maximum number of tokens that will be ever generated by the decentralized application, the tokens unlocked for distribution during the ICO, the tokens transferred to the founders or other participants, and the tokens locked for future use. The ICO plans are advertised by the company, mentioning the initial price, start date, end date, number of tokens unlocked, pricing schemes (constant pricing, incremental pricing, etc.). The token registry and the distribution rules are programmed as smart contracts and deployed on chain. Each buyer willing to invest in the decentralized application will need to acquire the necessary cryptocurrency and sign a transaction paying the requested sum in return for several tokens. The distribution contract will validate the deposited sum, and if all the rules hold, the token registry will update the buyer's account accordingly.

*Step C*. Development and application deployment. The decentralized application development will continue according to the roadmap presented in the whitepaper. Upon finalization, the application will be deployed on-chain, and all the users that acquired tokens during the ICO will be able to start using the decentralized application or will be able to sell the tokens to other interested parties.

*Step D*. Token Listing on public crypto exchanges. In order to facilitate the exchange of tokens between interested users, the company can list the token on one of the public crypto exchanges. Sellers can make their offers and Buyers can make their bids therefore, depending on the public interest in the launched business, the token can have the potential to raise its value. Being an off-chain exchange, only upon settlement the updated balance will be registered on chain mirroring the transaction that happened between the seller and the buyer. However, some of these exchanges chose to act as a custodian for the exchanged tokens and the actual registry is not updated on chain.

# 6. Conclusions

The distributed ledger technology has the potential of being a game changer in many domains, its recent developments being triggered not only by technology expectations but also by the social ones. Blockchains have enabled the development of decentralized applications but this process is still complicated due to the high availability of new and not yet mature technological solutions which address different DLT issues and offer new implementation opportunities. Blockchains are an effervescent innovation area therefore a review of new technological solutions, new architectures, design and implementation guidelines are needed to improve understanding and to ease the development of new applications.

In this paper, we provide a timely review of the existing DLT solutions that are categorized according to the 3-tiers of a conceptual architecture for decentralized applications defined by us. In this process we have considered references from both academia and private sector. For each identified solution we provide an overview description, by highlighting its applicability, the advantages and disadvantages with respect to other similar solutions. For each tier, a significative number of solutions have been evaluated in order to provide a holistic image over the DLT variations.

Finally, we provide a guideline for decentralized application development defining specific steps for decentralizing a system design and business model using DLT and for implementing and launching it as decentralized application. This is in line with nowadays efforts for investigating how the blockchain can be applied, integrated and used for decentralizing traditional systems such as medical systems, electricity grids, financial sector, etc. and for implementing new decentralized applications.

**Acknowledgements:** This work has been conducted within the eDREAM project Grant number 774478, co-funded by the European Commission as part of the H2020 Framework Programme (H2020-LCE-2017-SGS).